\documentclass[twoside]{dis04}
\def\runauthor{A.M.Cooper-Sarkar}
\def\shorttitle{PDFs extracted from ZEUS HERA-I data}
\usepackage{epsfig}

\newcommand{\leqsim}{\,\raisebox{-0.6ex}{$\buildrel < \over \sim$}\,}

\newcommand{\unit}[1] {\mbox{\hspace{0.3em}\rm #1}}

\newcommand{\asmz}{\alpha_s(M_Z^2)}
\newcommand{\msbar}{\mbox{$\overline{\rm{MS}}$}\ }
\def\etal{\mbox{\it et al.\ }}
\def\Journal#1#2#3#4{{#1} {\bf #2} {#3} ({#4})}

\def\NPB{{\it Nucl. Phys.} B}

\def\PLB{{\it Phys. Lett.}  B}

\def\PRD{{\it Phys. Rev.} D}

\def\EPJC{{\it Eur. Phys. J.} C}
\def\JPG{{\it J. Phys.} G.}

\def\JHEP{\it JHEP}

\begin{document}

\title{AN NLO QCD ANALYSIS OF INCLUSIVE CROSS-SECTION DATA AND JET PRODUCTION 
DATA FROM THE ZEUS EXPERIMENT AT HERA-I 
}

\author{A M COOPER-SARKAR}

\address{Denys Wilkinson Building\\Oxford University Physics Department\\
1 Keble Rd, Oxford OX1 3RH, UK\\
E-mail: a.cooper-sarkar@physics.ox.ac.uk }

\maketitle

\abstracts{The ZEUS inclusive differential cross-section data from HERA-I,    for charged and neutral
   current processes taken with $e^+$ and $e^-$ beams, 
   have been used in a new
   a NLO QCD analysis to extract 
   the parton distribution functions (PDFs) of the proton within a single 
   experiment. The precision of the
   high-$x$ gluon determination is improved by using 
   ZEUS differential cross-section data on inclusive
   jet production in $e^+ p$ scattering and di-jet production in $\gamma p$
   scattering.
}

\section{Introduction} 
ZEUS data from HERA-I on neutral and charged current (NC and CC) 
$e^+p$ and $e^-p$ 
cross-sections~\cite{NC967,CC947,NC989,CC989,NC9900,CC9900} have been used 
in a NLO QCD DGLAP
analysis in order to determine the 
parton distributions functions (PDFs) using data from within a single 
experiment. PDF determinations are usually global 
fits~\cite{mrst,cteq6,zeus2002}, which use fixed target 
DIS data as well as HERA data. In such analyses, the high statistics HERA NC 
$e^+p$ data, which span the range $6.3 \times 10^{-5} < x < 0.65,
2.7 < Q^2 < 30,000$GeV$^2$, 
have determined the low-$x$ sea and 
gluon distributions, whereas the fixed target data have determined 
the valence distributions and the higher-$x$ sea and gluon distributions. 
The most important inputs  
for determining the valence distributions have been the $\nu$-Fe  
and the $\mu D$  fixed target data, but these data sets suffer 
from uncertainties due to heavy target corrections~\cite{highq299}.
In the present analysis the ZEUS high $Q^2$ cross-section 
data~\cite{CC947,NC989,CC989,NC9900,CC9900} are used to determine the valence 
distributions thus eliminating such uncertainties. 
PDF fits to DIS data alone have suffered from a lack of information on the 
high-$x$ gluon. In the present analysis ZEUS jet production data are used to
constrain the gluon PDF.

The PDFs are presented with full accounting for uncertainties from correlated 
systematic errors (as well as from statistical and uncorrelated sources).
Peforming an analysis within a single experiment has considerable advantages
in this respect, since the global fits 
have found significant tensions between 
different data sets, which make a rigorous statistical treatment of 
uncertainties difficult.

\section{Analysis}

The kinematics
of lepton hadron scattering is described in terms of the variables $Q^2$, the
invariant mass of the exchanged vector boson, and Bjorken $x$, the fraction
of the momentum of the incoming nucleon taken by the struck quark (in the 
quark-parton model). The differential cross-sections for the NC and CC 
processes are 
given in terms of structure functions which are 
directly related to quark distributions. The $Q^2$ dependence of these 
structure functions is predicted by perturbative QCD. 
At $Q^2 \leqsim 1000$GeV$^2$ the NC structure function $F_2$ dominates the
charged lepton-hadron cross-section and for $x \leqsim 10^{-2}$, $F_2$ itself 
is sea quark dominated whereas its $Q^2$ evolution is controlled by
the gluon contribution, such that ZEUS data provide 
crucial information on quark and gluon distributions.
At high $Q^2$, the NC structure function $xF_3$ becomes increasingly 
important, 
and gives information on valence quark distributions. The CC interactions also
enable us to separate the flavour of the valence distributions 
at high-$x$. For a full explanation of the relationships between 
DIS cross-sections, 
structure functions, PDFs and the QCD improved parton model see for example
ref.~\cite{book}.
The present analysis is performed within the conventional
paradigm of leading twist, NLO QCD, with the
renormalisation and factorization scales chosen to be $Q^2$. 
The QCD predictions for the structure functions 
are obtained by solving the DGLAP evolution equations at NLO in 
the \msbar\ scheme. These equations yield the PDFs
 at all values of $Q^2$ provided they
are input as functions of $x$ at some input scale $Q^2_0$. 
The resulting PDFs are then convoluted with coefficient functions, to give the
structure functions which enter into the expressions for the cross-sections.
These coefficient functions are calculated using the general mass variable 
flavour number scheme of Roberts and Thorne~\cite{hq} for 
heavy quark production.

The PDFs for $u$ valence,  $d$ valence, 
total sea, and gluon 
are each parametrized  by the form 
\[
  p_1 x^{p_2} (1-x)^{p_3}( 1 + p_4 x)
\]
at $Q^2_0 = 7$GeV$^2$. No advantage in $\chi^2$ results from using more complex
polynomial forms. 
The normalisation parameters $p_1$ for the $d$ and $u$ valence and for the 
gluon are constrained to impose the number sum-rules and momentum sum-rule. 
The $p_2$ parameter which constrains the low-$x$ behaviour of the $u$ and $d$ 
valence distributions is set equal, 
since there is no information to constrain any difference. 
When fitting to ZEUS inclusive cross-section 
data alone it is necessary to constrain some of 
the parameters which control the high-$x$ sea and gluon shapes, 
because HERA-I data do not have high 
statistics at large-$x$, in the region where these distributions are small.   
There are two possible strategies towards making these constraints.

Firstly, the focus is on the valence distributions and
the number of free parameters describing the high-$x$ sea and gluon
is restricted. There are various ways to make restrictions. One possibility 
is to set $p_4=0$ for both these distributions. However this does not allow 
any structure at medium $x$  for $Q^2$ near the input scale.
Alternatively, the $p_4$ parameters can be freed but the $p_3$ parameters 
fixed to the values obtained in the ZEUS 
global fit~\cite{zeus2002}. In this case, model uncertainties
on the high-$x$ sea and gluon PDFs must include the effect of 
changing these fixed values of $p_3$ 
within the limits of their errors as determined in the global fit. 
In practice there is very little difference in the shapes and errors on the 
sea and gluon PDFs determined by these two ways of making restrictions. 
Distributions are presented for the latter choice. This fit has 10 free 
parameters and is called the ZEUS-O fit.

Secondly, a more ambitious strategy is pursued. ZEUS data on jet production are
included in the 
PDF fit. This not only gives more information on the high-$x$ gluon PDF, but 
also establishes that NLOQCD is able to simultaneously describe inclusive 
cross-sections and jet cross-sections providing a compelling test of QCD 
factorization. This fit is called the ZEUS-JETS fit. 
The method of incorporating the jet production data in the fit is discussed in
detail in Sec.~\ref{sec:jets}

For both analyses the strong coupling constant is fixed to $\asmz =  0.118$ 
and the following cuts are made on
the data: (i)~$W^2 > 20$GeV$^2$ to reduce the
sensitivity to target mass and higher 
twist contributions which become important at high $x$
and low $Q^2$; (ii)~$Q^2 > 2.5$GeV$^2$ to remain in the kinematic region where
perturbative QCD should be applicable. 
Full account has been taken of correlated experimental 
systematic errors by the Offset Method, 
as decribed in the previous ZEUS analysis~\cite{zeus2002} 
and discussed extensively
in~\cite{durham}. Further details of the analysis may be found in ~\cite{ICHEP}

\section{Results}
\subsection{ZEUS-O fit results}

\begin{figure}[tbp] 
\vspace*{13pt}
\centerline{
\psfig{figure=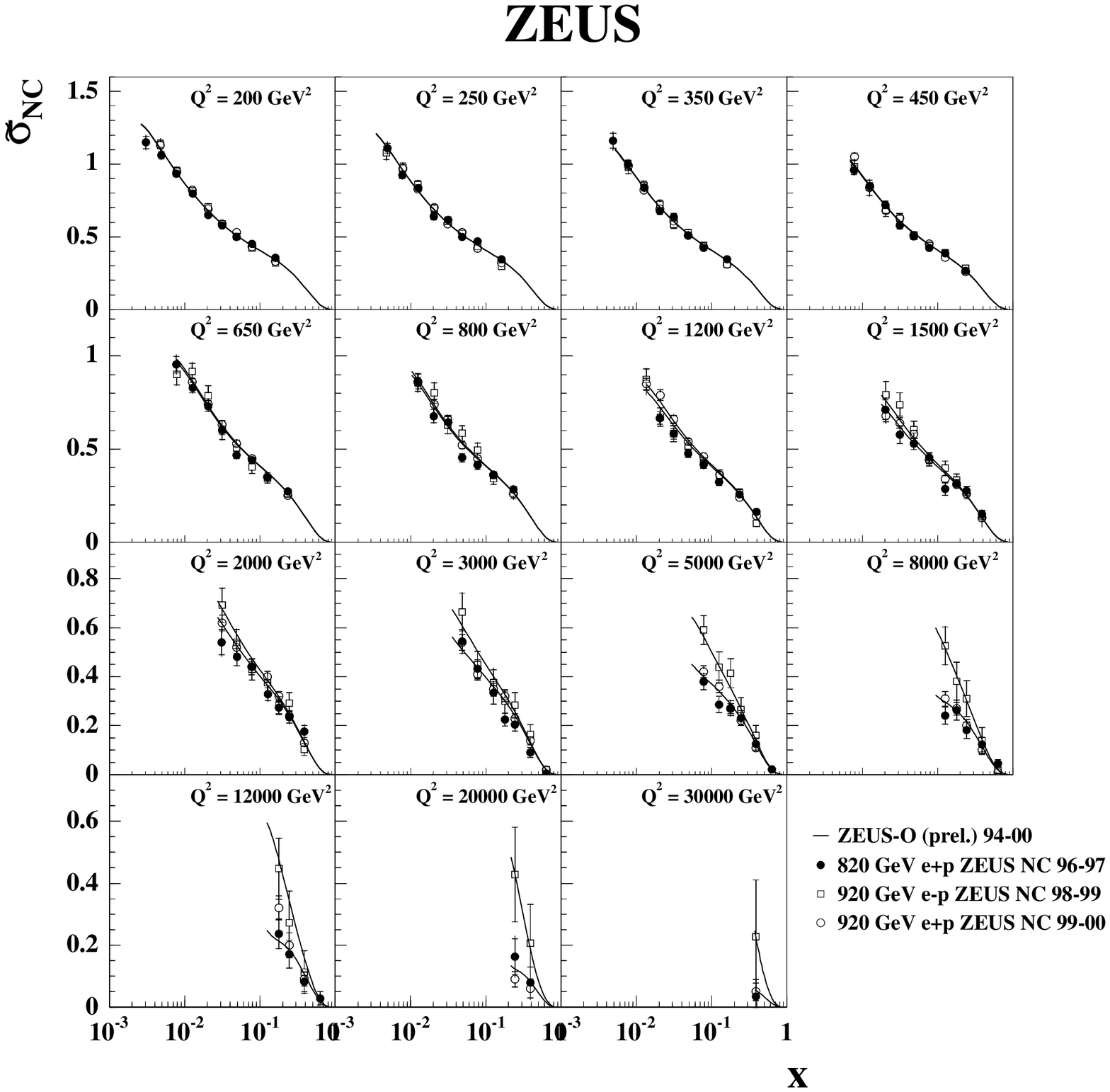,width=0.5\textwidth},
\psfig{figure=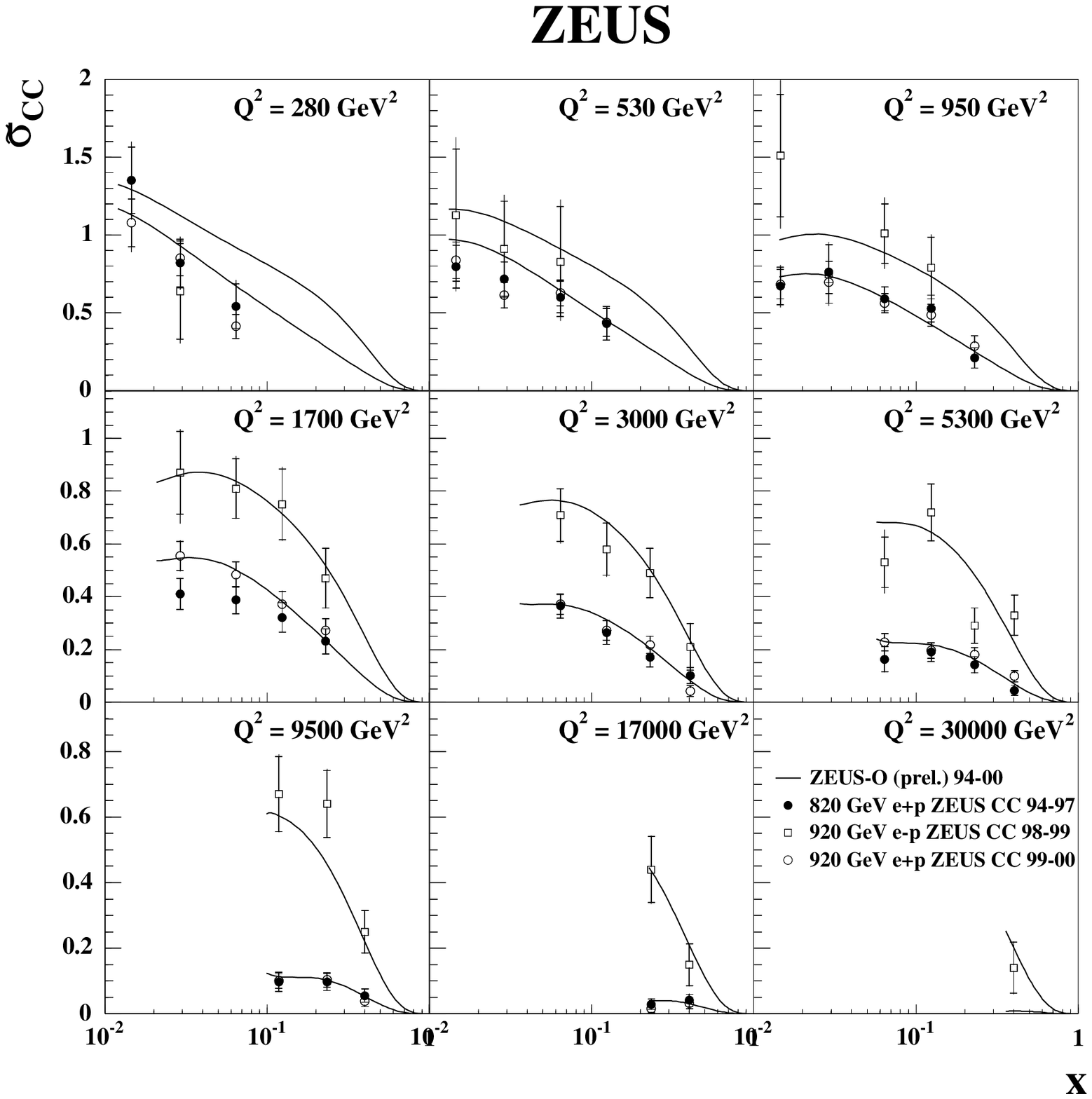,width=0.5\textwidth}}
\caption {Left plot: ZEUS-O fit to high $Q^2$ NC data.
Right plot: ZEUS-O fit to high $Q^2$ CC data.}
\label{fig:highQ2}
\end{figure}

For the ZEUS-O fit, inclusive cross-section data from 112 pb$^{-1}$ 
of HERA-I running are used. 
A good description of cross-section  
data over the whole range of $Q^2$ from 2.5 to $30000\unit{GeV}^2$
is obtained. The $\chi^2$ is 386 for 509 data points.
The quality of the fit to the high-$Q^2$ reduced cross-section 
data is illustrated in Fig.~\ref{fig:highQ2}. 

\begin{figure}[tbp] 
\vspace*{13pt}
\centerline{
\psfig{figure=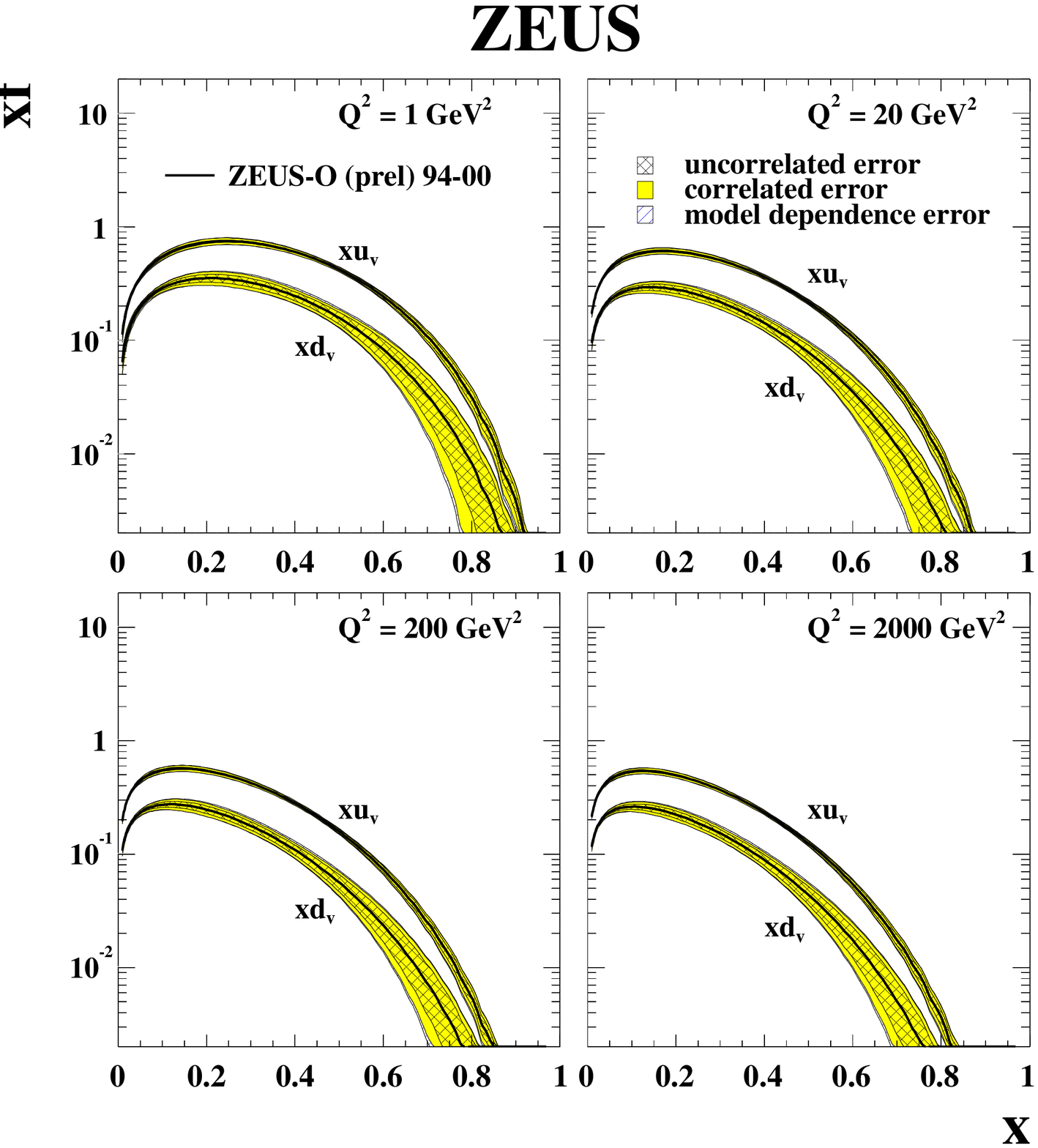,width=0.5\textwidth},
\psfig{figure=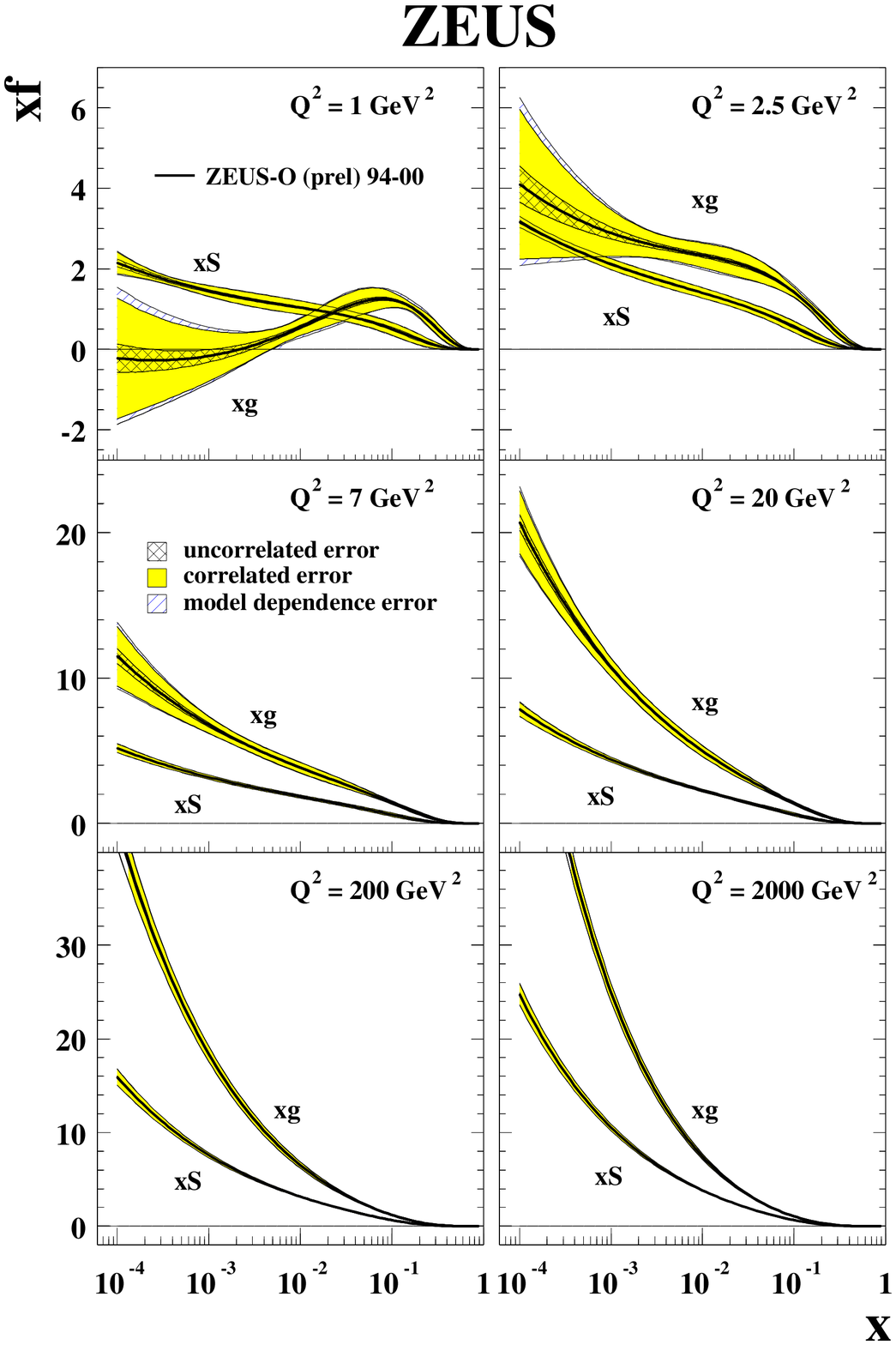,width=0.5\textwidth}}
\caption {Left plot: Valence PDFs extracted from the ZEUS-O fit.
Right plot: Gluon and sea PDFs extracted from the ZEUS-O fit.
}
\label{fig:PDFSA}
\end{figure}

The valence distributions are shown in Fig~\ref{fig:PDFSA}. 
Although the high-$x$ valence 
distributions are not quite as well constrained as they are in global fits 
including fixed target data~\cite{zeus2002,mrst,cteq6} they are becoming 
competitive, particularly for the less well known $d$-valence distribution. 
Furthermore, they are free from uncertain heavy target corrections.

The gluon and sea distributions are shown in Fig.~\ref{fig:PDFSA}. They 
exhibit the familiar features of the sea rising at low-$x$ down to surprisingly
 low $Q^2$ values, and the gluon flattening and then becoming valence-like 
in the same kinematic region. They are as well determined as the corresponding
distributions of the global 
fits~\cite{zeus2002,mrst,cteq6} at low-$x$, since the ZEUS data were crucial
in determining these distributions for all the fits. 
At high-$x$ they have the same uncertainties as the global fits because of the 
constraints that have been applied. 
 
The experimental errors represent the most significant source of uncertainty
on these distributions. 
Variation of analysis choices, such as the value of $Q^2_0$, 
the minimum $Q^2$ of data entering the fit, and changing 
the form of the parametrization at $Q^2_0$,  
do not produce a large model uncertainty. 

\subsection{ZEUS-JETS fit results}
\label{sec:jets}

The gluon PDF contributes only indirectly to the 
inclusive DIS cross-sections. However it  makes a direct contribution to
jet cross-sections through the Boson-Gluon-Fusion process. In the present 
analysis the mid to high-$x$ 
gluon distribution is constrained by including in the fit 
two ZEUS measured jet cross sections:
\begin{itemize}
\item 30 data points from six  DIS inclusive 
jet differential cross-sections as a function of the transverse energy, 
$E_T^B$, in the Breit frame, for different $Q^2$ bins~\cite{DISjets}.
\item 38 data points from six direct photoproduction di-jet 
cross-sections as a function of the transverse energy of the most energetic 
jet, $E_T^{jet1}$ in the lab, for different jet rapidity ranges~\cite{phojets}.
\end{itemize}
For each of these data sets the calorimeter energy scale uncertainty 
and the normalization uncertainty have been treated as 
correlated systematic errors.

The programme of Frixione and Ridolfi~\cite{FrixRid} is used to compute 
NLO QCD cross sections for photoproduced di-jets and DISENT~\cite{disent}
is used to compute NLO QCD cross sections for jet production in DIS.
 These programmes are too slow to be used every iteration of a fit. 
Thus these codes are used to produce grids,in 
$\xi$ (the parton momentum fraction) and $\mu_F^2$ (the factorization scale),
of weights which represent sub-process cross-sections  
for each flavour of parton 
(gluon, up-type, down-type). This is done for each cross-section bin 
and these weights are then used to reconstruct the cross-sections 
as follows
\begin{equation}
 d\sigma(jets) = \sum_{a=u,d,g} \int\int d\xi d\mu^2_F f_a(\xi,\mu^2_F) 
\sigma_a(\xi,\mu_F^2,\alpha_s(\mu_R)) 
\end{equation}
where $f_a$ is the PDF for parton type $a$ at parton momentum fraction
$\xi$ and scale $\mu_F$, and
$\sigma_a$ is the sub cross-section weight. 
 The factorization scale is chosen as
$\mu_F= Q$ for the DIS jets, and the renormalization scale is $\mu_R=E_T$ (with
$\mu_R=Q$ as a cross-check). For the photoproduced di-jets the standard 
scale choices are $\mu_R= \mu_F = E_T/2$  (where $E_T$  is the 
summed $E_T$ of final state partons). The grids reproduce the NLO predictions
 to better than $0.5\%$. The predictions are then multiplied by hadronization
 corrections and $Z_O$ corrections before they are fitted to data.

The predictions for photoproduced jets can obviously be 
influenced by the choice of the input photon PDF. Thus the analysis is 
restricted to direct photoproduction, defined such 
that all data points satisfy 
$x_\gamma^{OBS} > 0.75$, where $x_\gamma^{OBS}$ 
is a measure of the fraction of the 
photon's momentum which enters into the hard scattering, see 
ref.~\cite{phojets}. This minimizes sensitivity to the choice of the 
photon PDF. The AFG photon PDF is used to make the standard predictions, 
and the GRV photon PDF is
used as a cross-check. There is no visible difference to the 
extracted proton PDFs.

\begin{figure}[tbp] 
\vspace*{13pt}
\centerline{
\psfig{figure=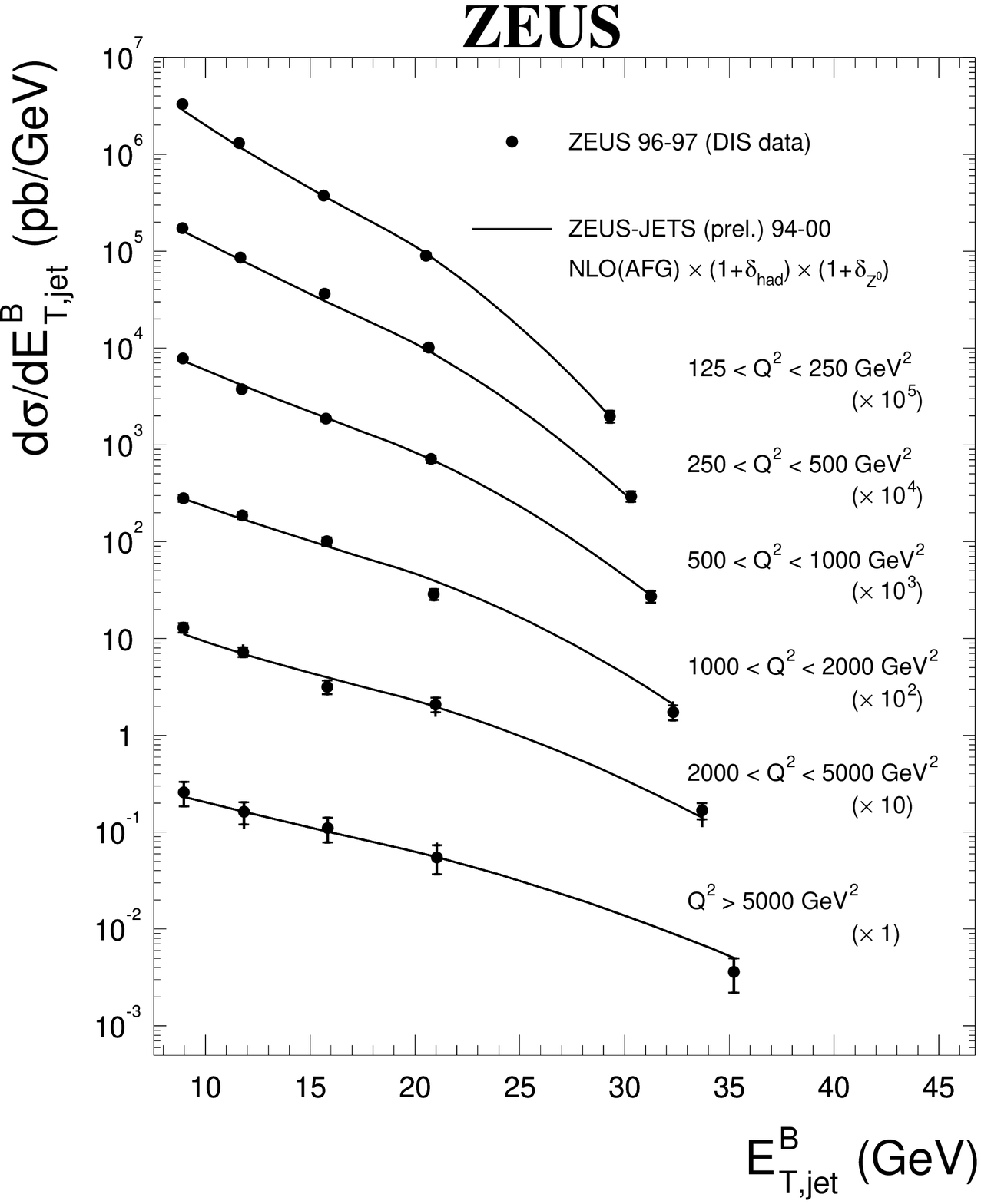,width=0.5\textwidth},
\psfig{figure=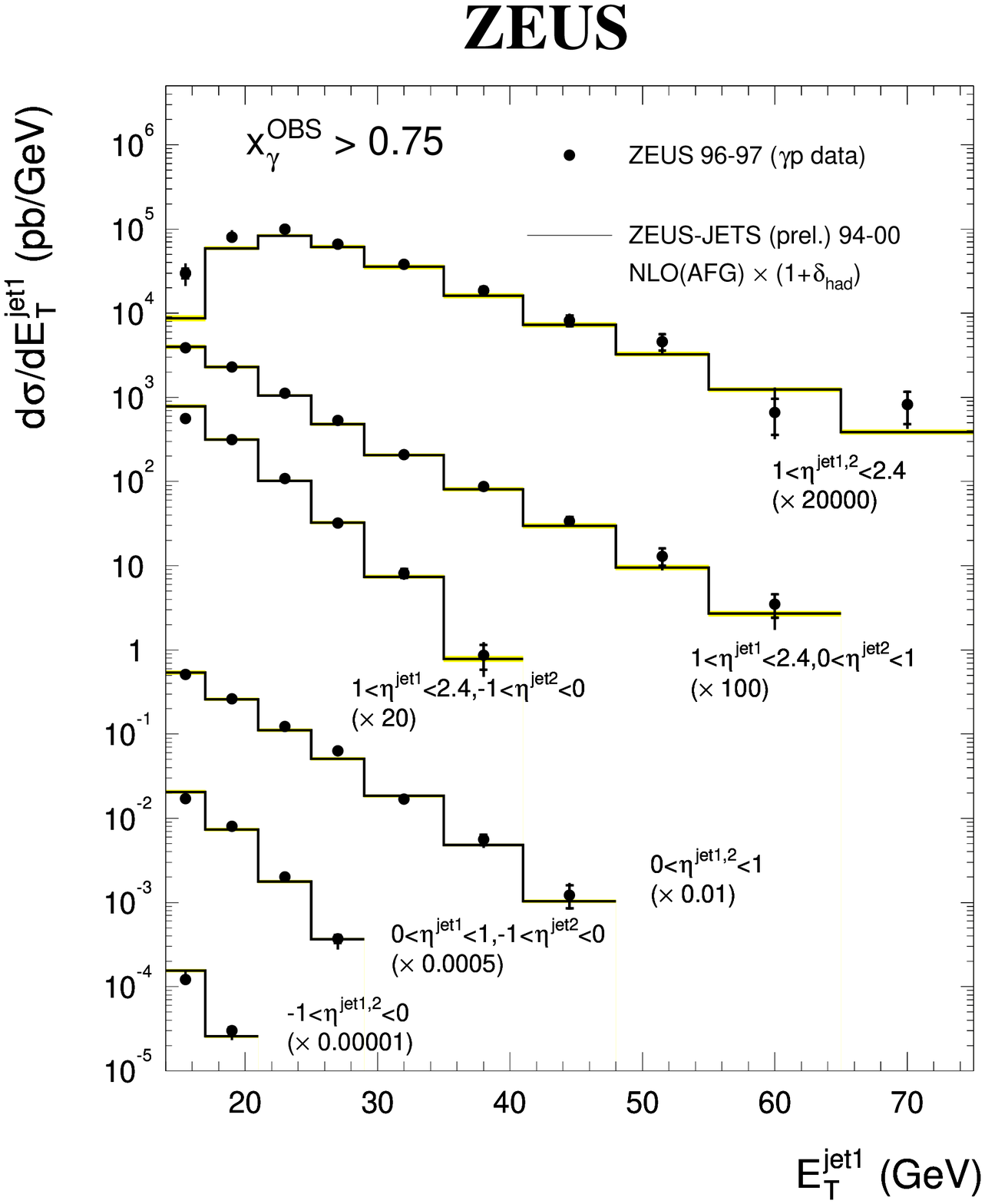,width=0.5\textwidth}}
\caption {Left plot: ZEUS-JETS fit to DIS jet data.
Right plot: ZEUS-JETS fit to direct photo-produced dijet data.}
\label{fig:jets}
\end{figure}

The jet data are input to fits in which the $p_3$ and the 
$p_4$ parameters, which control the high-$x$ behaviour of 
the sea and the gluon PDF, are freed. It is found that 
these data constrain the gluon distribution in the range $ 0.01 < x < 0.1$.
However, the sea distribution is not significantly constrained, so that  
the strategy of fixing one of the high-$x$ sea parameters is retained. 
Thus the ZEUS-JETS fit has 11 free parameters. The fit $\chi^2$ is 479 for 
577 data points. The quality of the fit to the jet data is illustrated in 
Fig~\ref{fig:jets}. 

\begin{figure}[tbp] 
\vspace*{13pt}
\centerline{
\psfig{figure=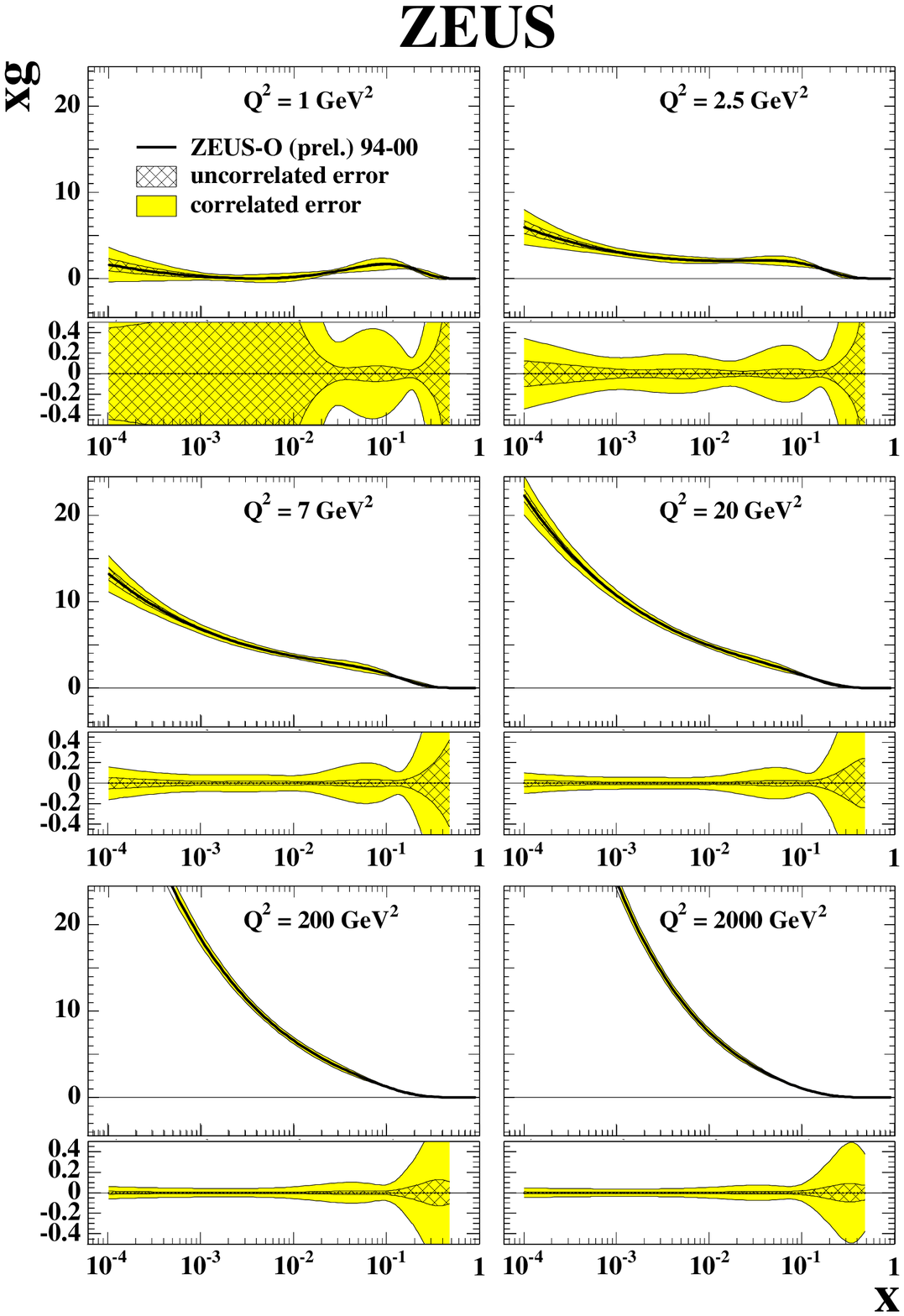
,width=0.5\textwidth},
\psfig{figure=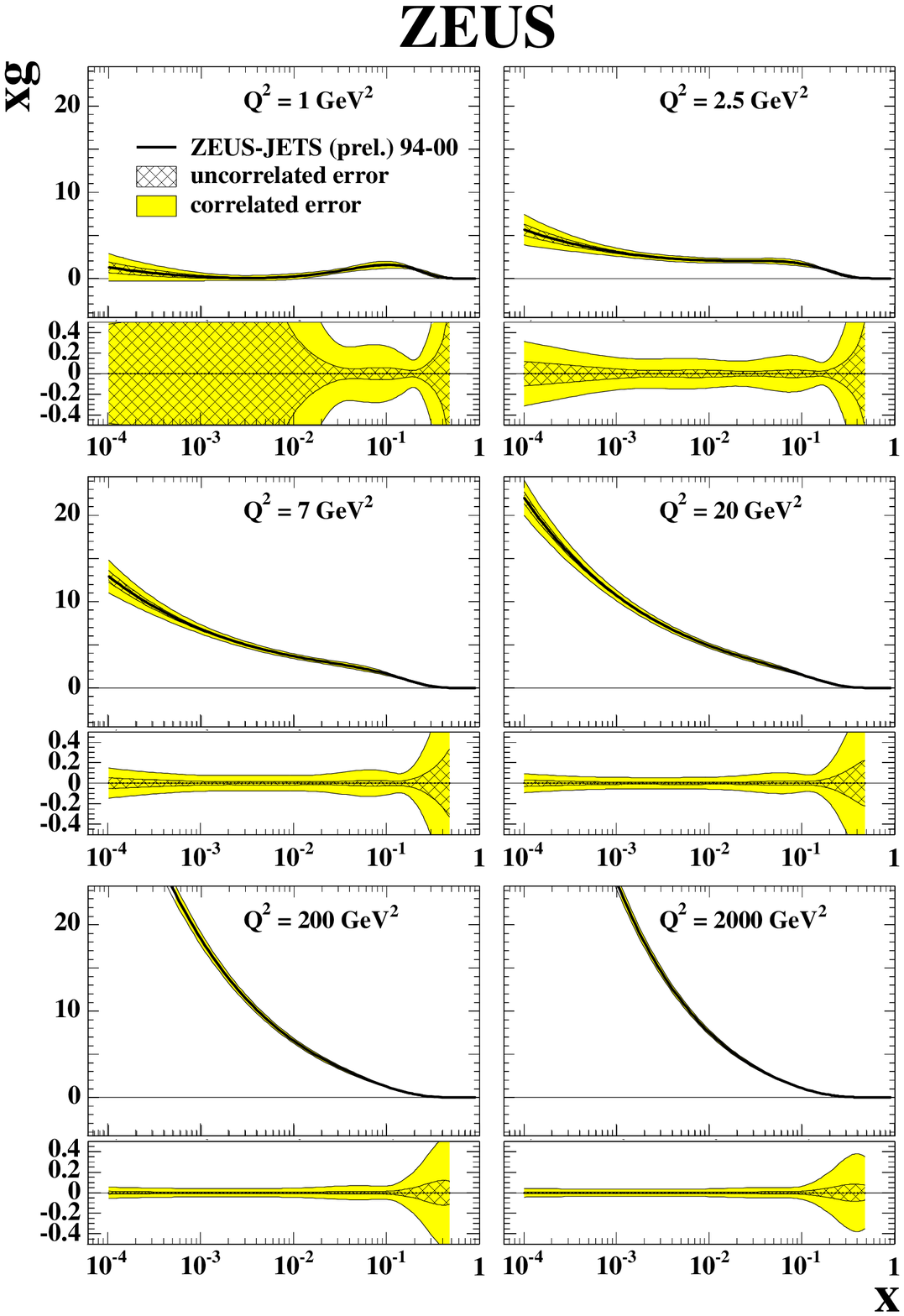
,width=0.5\textwidth}}
\caption {Left plot: Gluon distributions extracted from an 11-parameter PDF 
fit excluding jet production data.
Right plot: Gluon distributions extracted from the an 11-parameter PDF fit
including jet production data (ZEUS-JETS fit). The uncertainties on these 
distributions are shown beneath each distribution as fractional differences 
from the central value. Note that model uncertainty is not visible for 
these fits}
\label{glujets}
\end{figure}

Fig.~\ref{glujets} compares the gluon distribution and its errors for 11
parameter fits
including and not including the jet data.  Although the jet 
data constrain the gluon directly only in the range  
$ 0.01 \leqsim x \leqsim 0.1$, the momentum sum-rule ensures that the indirect 
constraint of this data is still significant at higher $x$. 
The decrease in the uncertainty on 
the gluon distribution is striking, even up to high $Q^2$.

\begin{figure}[tbp] 
\vspace*{13pt}
\centerline{
\psfig{figure=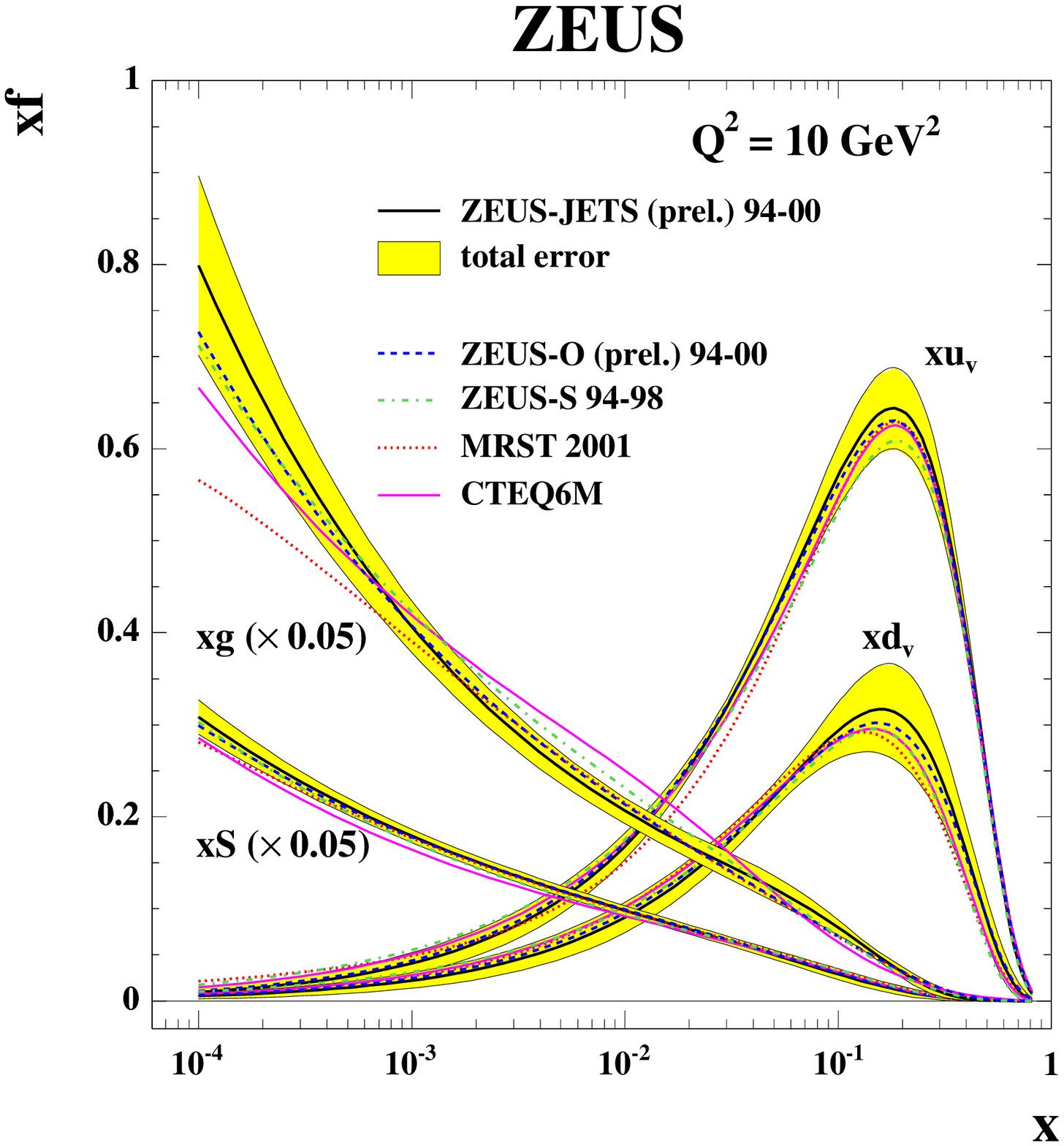,width=0.5\textwidth}
}
\caption {
PDFs extracted from the new ZEUS-O and  ZEUS-JETS fits compared to the 
previous ZEUS-S analysis and the MRST and CTEQ PDFs.
}
\label{fig:summary}
\end{figure}

Fig.~\ref{fig:summary} (left hand plot) 
compares all the PDFs for ZEUS-O and ZEUS-JETS 
fits, illustrating that the valence and sea PDFs are not significantly 
influenced by the input of the jet data. The PDFs from the previous ZEUS-S 
global fits~\cite{zeus2002} and the MRST and CTEQ PDFs~\cite{mrst,cteq6} 
are also shown on this figure. There is good agreement between all the ZEUS
PDF extractions. The CTEQ and MRST PDFs are also compatible considering the 
size of the PDF uncertainties.

\end{document}